# Strict Authentication Watermarking with JPEG Compression (SAW-JPEG) for Medical Images

**Jasni Mohamad Zain**
*Faculty of Computer Systems & Software Engineering, Universiti Malaysia Pahang, Lebuhraya Tun Razak, 26300, Gambang, Kuantan, Pahang, Malaysia.*
E-mail: jasni@ump.edu.my
Tel: +609 5492133; Fax: +609 5492144

**Abstract**

This paper proposes a strict authentication watermarking for medical images. In this scheme, we define region of interest (ROI) by taking the smallest rectangle around an image. The watermark is generated from hashing the area of interest. The embedding region is considered to be outside the region of interest as to preserve the area from distortion as a result from watermarking. The strict authentication watermarking is robust to some degree of JPEG compression (SAW-JPEG). JPEG compression will be reviewed. To embed a watermark in the spatial domain, we have to make sure that the embedded watermark will survive JPEG quantization process. The watermarking scheme, including data embedding, extracting and verifying procedure were presented. Experimental results showed that such a scheme could embed and extract the watermark at a high compression rate. The watermark is robust to a high compression rate up to 90.6%. The JPEG image quality threshold is 60 for the least significant bit embedding. The image quality threshold is increased to 61 for 2nd and 3rd LSB manipulations.

**Keywords:** Strict authentication, watermarking, JPEG compression, region of interest

## 1. Introduction

Image compression seeks to reduce the number of bits required to represent the image information. Two fundamental properties used in image compression are removal of redundancy and reduction of irrelevant content. Irrelevant content may include information not perceived by the viewer, namely the human visual system (HVS). Three types of redundancy may be exploited; spatial redundancy or correlation between neighbouring pixels; spectral redundancy or correlation between different frequency bands; and temporal redundancy or correlation between adjacent frames in a sequence of images (in video applications).

Compression algorithms can be divided into two main groups, lossless and lossy methods. In lossless compression schemes, only the redundancy is exploited, and the image is recorded in a more efficient manner. All the information is retained and so the reconstructed image is numerically identical to the original image. In lossy compression, information deemed irrelevant to the visual perception of the human viewer is discarded and so the compressed image cannot be perfectly reconstructed and distortion is introduced into the reconstructed image.

While lossless compression does not harm a watermarking system in any way (the original data can be perfectly reconstructed), lossy compression methods introduce distortion that has to be taken into account in watermarking applications. Lossy compression techniques are nowadays being commonly used as a means to effect a reduction on the requirement for bandwidth and storage space. It is therefore necessary to study the effects of lossy image compression on watermarking systems.

It should be observed that the design goal of lossy compression systems is opposed to that of watermark embedding systems. The HVS model of the compression system attempts to identify and discard perceptually insignificant information of the image, whereas the goal of the watermarking system is to embed the watermark information without altering the visual perception of the image. An optimal compression or de-noising system would immediately discard any such watermark information. Fortunately, all current compression methods are not optimal and allow watermarking schemes to be devised that will embed watermark information that is robust.

It remains unresolved how lossy compression should best be employed for the storage and transmission of medical images. There is little guidance from the scientific literature, professional practice standards, regulatory authorities, or the common law. Although lossy compression schemes are included in medical standards such as DICOM, their clinical use is not defined; it is only that the technology is available for use at the discretion of the user or implementer.

There is no good metric by which to judge lossy compression schemes or determine appropriate threshold levels for diagnostic use. Quantitative metrics based on an analysis of the image pixels such as Mean Squared Error (MSE) and Peak Signal-to-Noise Ratio (PSNR) do not correlate well with observers' opinions of image quality, or the measurement of observers' performance when undertaking diagnosis. Metrics based on models of human visual perception are still in their infancy. They have not been thoroughly compared to observer performance for medical applications (Clunie 2000).

Hybrid lossless/lossy compression schemes have been developed for medical applications. These identify regions of images that are determined by some criterion to be of little or no clinical interest. These regions are then either discarded or compressed with greater loss. The remaining regions, which contain the regions of clinical interest, are compressed using a lossless compression scheme. This approach can result in a high compression overall and retain the effective quality of a lossless compression scheme. The difficulty is to determine the areas of clinical interest. There has been work to find automate algorithms, but the only reliable method has been to determine regions defined by physical characteristics Some early CT compression schemes did not encode information outside the circular reconstructed area at all (perimeter coding) and were very effective. However, if such areas are filled with a constant pixel value then most general-purpose lossless image compression schemes perform equally well.

Nowadays, there exist watermarking methods for virtually every kind of digital media: text documents (Su et al. 1998, Brassil et al. 1999), images (Tsai et al. 2004, Zhang et al. 2003, Paquet et al. 2003), video (Sun and Chang 2003, Okada et al. 2002), audio (Li and Xue 2003, Yan et al. 2004), even for 3D polygonal models (Kwon et al. 2003, Benedens and Busch 2000), maps (Barni et al. 2001) and computer programs (Monden et al. 2000).

## 2. Previous Research

JPEG (Wallace 1991) is currently the most frequently used compression algorithm for medical imaging. For example it is included within the DICOM standard. Improved compression algorithms such as JPEG2000, will replace JPEG in time. For the purposes of this work, the watermarking method will focus specifically on JPEG, although the method should be extensible to other compression schemes based on a block compression scheme.

In this section, we briefly review the JPEG lossy compression standard (Wallace 1991). At the input to the JPEG encoder, the source image, X, is grouped into ρ nonoverlapping 8x8 blocks, $X_p$. Each block is sent sequentially to the Discrete Cosine Transform (DCT). Instead of representing each 8x8 matrix, we can rewrite it as a 64x1 vector following the "zigzag" order (Wallace 1991). Therefore the DCT coefficients, $F_p$, of the vector, $X_p$, can be considered as a linear transformation of $X_p$ with a 64x64 transformation matrix D, such that,

$$F_p = DX_p \quad \text{(Equation 1)}$$

The two-dimensional DCT of an M x N image X is defined as follows:

$$B_{pq} = \alpha_p \alpha_q \sum_{m=0}^{M-1}\sum_{n=0}^{N-1} X_{mn} \cos\frac{\pi(2m+1)p}{2M}\cos\frac{\pi(2n+1)q}{2N}, \quad \begin{array}{l} 0 \le p \le M-1 \\ 0 \le q \le N-1 \end{array}$$

$$\alpha_p = \begin{cases} 1/\sqrt{M}, & p = 0 \\ \sqrt{2/M}, & 1 \leq p \leq M \end{cases} \quad \alpha_q = \begin{cases} 1/\sqrt{N}, & q = 0 \\ \sqrt{2/N}, & 1 \leq q \leq N \end{cases} \quad \text{(Equation 2)}$$

The values $B_{pq}$ are called the DCT coefficients of X. The DCT is an invertible transform. Each of the 64 DCT coefficients is uniformly quantized with a 64-element quantization table, Q.

**Table 1:** JPEG quantization table

| 16 | 11 | 10 | 16 | 24  | 40  | 51  | 61  |
|----|----|----|----|-----|-----|-----|-----|
| 12 | 12 | 14 | 19 | 26  | 58  | 60  | 55  |
| 14 | 13 | 16 | 24 | 40  | 57  | 69  | 56  |
| 14 | 17 | 22 | 29 | 51  | 87  | 80  | 62  |
| 18 | 22 | 37 | 56 | 68  | 109 | 103 | 77  |
| 24 | 35 | 55 | 64 | 81  | 104 | 113 | 92  |
| 49 | 64 | 78 | 87 | 103 | 121 | 120 | 101 |
| 72 | 92 | 95 | 98 | 112 | 100 | 103 | 99  |

In JPEG, the same table is used on all blocks of an image. Quantization is defined as the division of each DCT coefficient by its corresponding quantizer step size, and rounding to the nearest integer:

$$\tilde{f}_p(v) \equiv IntegerRound\left(\frac{F_p(v)}{Q(v)}\right) \quad \text{(Equation 3)}$$

where v = 1…64. In Equation 3, is the output of the quantizer. We define , a quantized approximation of Fp, as

$$\tilde{F}_p \equiv \tilde{f}_p(v) \cdot Q(v) \quad \text{(Equation 4)}$$

In addition to quantization, JPEG also includes scan order conversion DC differential encoding, and entropy coding.

Inverse DCT (IDCT) is used to convert to the spatial domain image block

$$\tilde{X}_p = D^{-1}\tilde{F}_p \quad \text{(Equation 5)}$$

All blocks are then tiled to form a decoded image frame. Theoretically, the results of IDCT are real numbers. However the brightness of an image is usually represented by an 8-bit integer from 0 to 255 and thus a rounding process mapping those real numbers to integers is necessary.

**Figure 1:** JPEG Encoder and Decoder

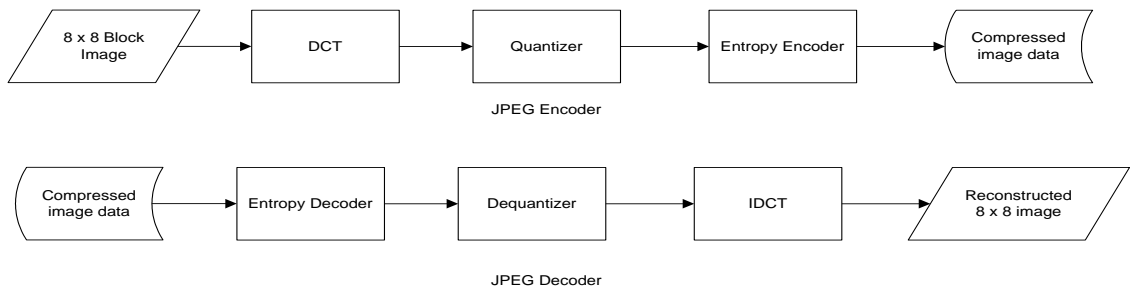

## 3. Research Method

### 3.1 Watermark

The watermark is generated by creating a hash value from the region of interest (inside the rectangle), *X* of size *m x n*. The pixels will be arranged in a string, *S*.

$$S = B(X_{(1,1)} X_{(1,2)} ... X_{(1,m)} X_{(2,1)} ... X_{(m,n)}),\quad \text{(Equation 6)}$$

where $X_{mn}$ is the 8 bit binary value of each pixel.
The hash value is obtained by applying a hash function to the string

$$Hash = H(S) \quad \text{(Equation 7)}$$

where H is any hash function such as MD5 and SHA256.

## 3.2 Embedding Region and Domain

**Figure 2:** Embedding Region

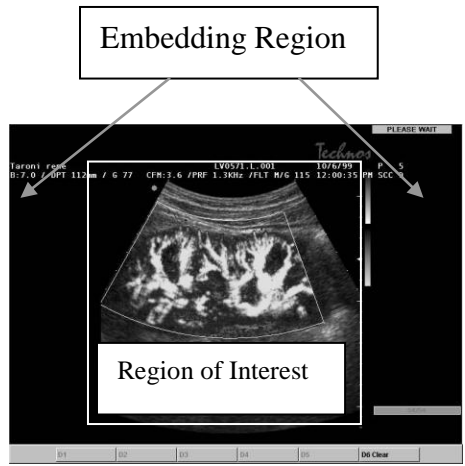

The embedding region is considered to be outside the region of interest in order to prevent distortion to the area as a result of adding the watermark. In an ultrasound image, the embedding region is normally a dark region with pixel values 0. This feature will be exploited to create a reversible or invertible watermarking.

In strict authentication watermarking, it is vital that the system will detect any change to the image. Fragile watermarking is the most appropriate as any change in the image will also affect the watermark. Least Significant Bit (LSB) watermarking has an advantage as the method of choice, as it is well known that LSB is vulnerable and easy to manipulate.

## 3.3 Security

A watermark is secure if it is able to resist intentional tampering by an attacker. This would include remaining secure even when the attacker knows the algorithm for embedding and extracting the watermark.

The strength of the security of the watermark will depend on the key chosen. A typical attack would involve removing the watermark, changing the image, then recalculating and embedding the new hash value into the embedding area. If the key for calculating the hash value remains secret, then the system may be considered secure. The secret key can be used to create the hash value and to create a random embedding. These will be examined in turn.

A key can be used to create the hash for the selected region. In this method, the sender and recipient will use the same key to carry out the hash function. The hash value obtained will be used as the watermark. At the recipient end, the key will be used to carry out the hash function on the received image and the hash value will be compared with the hash extracted.

**Figure 3**: Key for hash

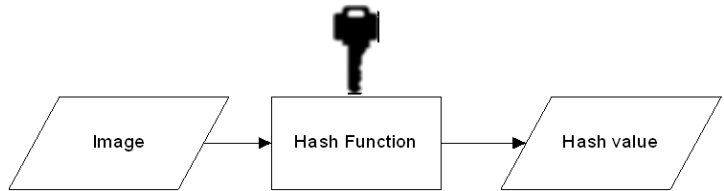

A key used for embedding will determine the random mapping of watermark values into the embedding region as in figure 4.

**Figure 4**: Hash value mapping in the embedding region

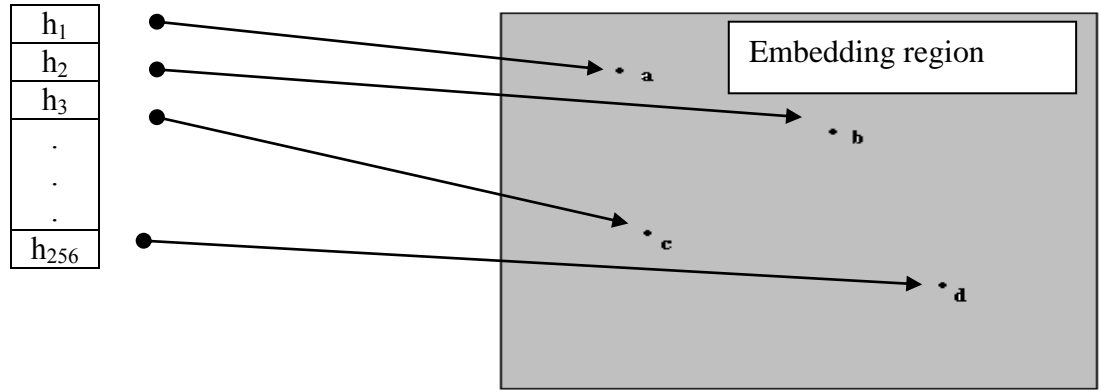

This supposes that the number of points or pixels in the embedding region is greater than or equal to the number of bits in the hash value holds. As an example, suppose the pixels are arranged as a simple raster scan as in figure 5.

**Figure 5**: Embedding region of 5 x 4 pixels

| 1 | 2 | 3 | 4 | 5 |
|---|---|---|---|---|
| 6 | 7 | 8 | 9 | 10 |
| 11 | 12 | 13 | 14 | 15 |
| 16 | 17 | 18 | 19 | 20 |

which may be described by the mapping function of equation 8:

$$f(x) = x \bmod n \qquad \text{(Equation 8)}$$

where x is the bit position and $x \in \{1, h\}$ and $n$ is the total number of pixels available for embedding. In this example, we use h=20 to make full use of the embedding region. Applying equation 4.3, bit position one will be located in pixel number one, bit position two will be located in pixel number 2 and so on. By using a key, $k$, the position will be randomised. If a simple function, e.g. equation 4.4 is applied,

$$f(x) = kx \bmod n \qquad \text{(Equation 9)}$$

where k is a prime key, then the mapping will be a randomised one-to-one mapping. This is illustrated for k=37 and n=20 in table 2.

**Table 2:** Mapping for k=37, n=20

| x | 1 | 2 | 3 | 4 | 5 | 6 | 7 | 8 | 9 | 10 | 11 | 12 | 13 | 14 | 15 | 16 | 17 | 18 | 19 | 20 |
|---|---|---|---|---|---|---|---|---|---|---|---|---|---|---|---|---|---|---|---|---|
| **F(x)** | 18 | 15 | 12 | 9 | 6 | 3 | 20 | 17 | 14 | 11 | 8 | 5 | 2 | 19 | 16 | 13 | 10 | 7 | 4 | 1 |

The method may be extended so that a number, h, of hash values are distributed within a region having many more pixel points, n so that the results appears as a sparse random distribution. The method is illustrated for k=37, h= 20 and n=100 in table 3.

**Table 3:** Mapping for k=37, h=20, n=100

| x    | 1  | 2  | 3  | 4  | 5  | 6  | 7  | 8  | 9  | 10 | 11 | 12 | 13 | 14 | 15 | 16 | 17 | 18 | 19 | 20 |
|------|----|----|----|----|----|----|----|----|----|----|----|----|----|----|----|----|----|----|----|----|
| f(x) | 38 | 75 | 12 | 49 | 86 | 23 | 60 | 97 | 34 | 71 | 8  | 45 | 82 | 19 | 56 | 93 | 30 | 67 | 4  | 41 |

If the embedding region is 10 x 10 pixels, then the distribution of embedding will be pictured as in figure 6.

**Figure 6**: Distribution of embedding for k=37, h=20, n=100

|    |    |    | 19 |    |    |    | 11 |    |    |
|----|----|----|----|----|----|----|----|----|----|
|    | 3  |    |    |    |    |    |    | 14 |    |
|    |    | 6  |    |    |    |    |    |    | 17 |
|    |    |    | 9  |    |    |    | 1  |    |    |
| 20 |    |    |    | 12 |    |    |    | 4  |    |
|    |    |    |    |    | 15 |    |    |    | 7  |
|    |    |    |    |    |    | 18 |    |    |    |
| 10 |    |    |    | 2  |    |    |    |    |    |
|    |    | 13 |    |    |    | 5  |    |    |    |
|    |    |    | 16 |    |    | 8  |    |    |    |

This simple method relies on the use of symmetric keys, which has an associated problem of key management. This is beyond the scope of this research. In practice asymmetric key systems are favoured; these are discussed in the next section.

### 3.4  Hashing – SHA256

The Secure hash Algorithm (SHA) was developed by the National Institute of Standards and Technology (NIST) and published as a federal information processing standard (FIPS PUB 180) in 1990. The algorithm is an iterative, one-way hash function that can process a message to produce a condensed representation called a *message digest*. The algorithm enables the integrity of a message to be determined and any change to the message will, with a very high probability, result in a different message digest. This property is useful in the generation and verification of digital signatures and message authentication codes. It is based on a public/private key, and thus overcomes the problem of key management.

To embed a watermark in the spatial domain, it is necessary to ensure that the embedded watermark will survive JPEG quantization process. JPEG processes images in 8 x 8 blocks, and so the method in which the watermark is embedded should be based on this same block structure. The process may be illustrated by encoding an 8 x 8 sub-image using JPEG. Consider if a '1' is embedded into the whole of the LSB plane of the 8 x 8 block as depicted by figure 7.

**Figure 7**: '1' bit embedded in 8x8 block

| 1 | 1 | 1 | 1 | 1 | 1 | 1 | 1 |
|---|---|---|---|---|---|---|---|
| 1 | 1 | 1 | 1 | 1 | 1 | 1 | 1 |
| 1 | 1 | 1 | 1 | 1 | 1 | 1 | 1 |
| 1 | 1 | 1 | 1 | 1 | 1 | 1 | 1 |
| 1 | 1 | 1 | 1 | 1 | 1 | 1 | 1 |
| 1 | 1 | 1 | 1 | 1 | 1 | 1 | 1 |
| 1 | 1 | 1 | 1 | 1 | 1 | 1 | 1 |
| 1 | 1 | 1 | 1 | 1 | 1 | 1 | 1 |

After the DCT transform of the block, figure 8 is the result.

**Figure 8**: DCT Transform of figure 4.14

| 8 | 0 | 0 | 0 | 0 | 0 | 0 | 0 |
|---|---|---|---|---|---|---|---|
| 0 | 0 | 0 | 0 | 0 | 0 | 0 | 0 |
| 0 | 0 | 0 | 0 | 0 | 0 | 0 | 0 |
| 0 | 0 | 0 | 0 | 0 | 0 | 0 | 0 |
| 0 | 0 | 0 | 0 | 0 | 0 | 0 | 0 |
| 0 | 0 | 0 | 0 | 0 | 0 | 0 | 0 |
| 0 | 0 | 0 | 0 | 0 | 0 | 0 | 0 |
| 0 | 0 | 0 | 0 | 0 | 0 | 0 | 0 |

To survive the quantization process, the value must be preserved through transformation and inverse transformation, that is

$$F_p = \tilde{F}_p \qquad (4.10)$$

To achieve this, $\dfrac{F_p(v)}{Q(v)}$ must be the integer and have no effect on the rounding process. In particular the DC quantization coefficient should be equal to the dc component in order to preserve an integer result, and all other quantization coefficients should be scaled accordingly. For higher compression rate, to preserve an integer value, the embedded level must be increased, which will naturally have an effect on the quality of the image.

By designing the watermark embedding algorithm around the properties of the compression scheme, it is possible to preserve the watermark values. In this case, a priori knowledge of the quantization algorithm allows the DC coefficient to be unchanged through the compression/decompression process.

The complete process is shown in figure 9 and comprises of the following steps:
1) Define area: This will define the Region of interest (ROI) where the smallest rectangle is obtained. Please refer to section 3.2.
2) SHA256: refer to section 3.2.
3) Embedding: Embed the hash value in the Region of Non-Interest (RONI) in the LSB. Since JPEG uses 8x8 blocks, we try to embed 1-bit in an 8x8 block. We only need 256 8x8 blocks to be able to embed the hash value.

4) JPEG Compression: Compression is performed on the watermarked image.
5) Extraction: The watermark is recovered from the watermarking area.
6) Authentication: The original hash and the extracted hash value are compared.

**Figure 9**: Watermarking scheme

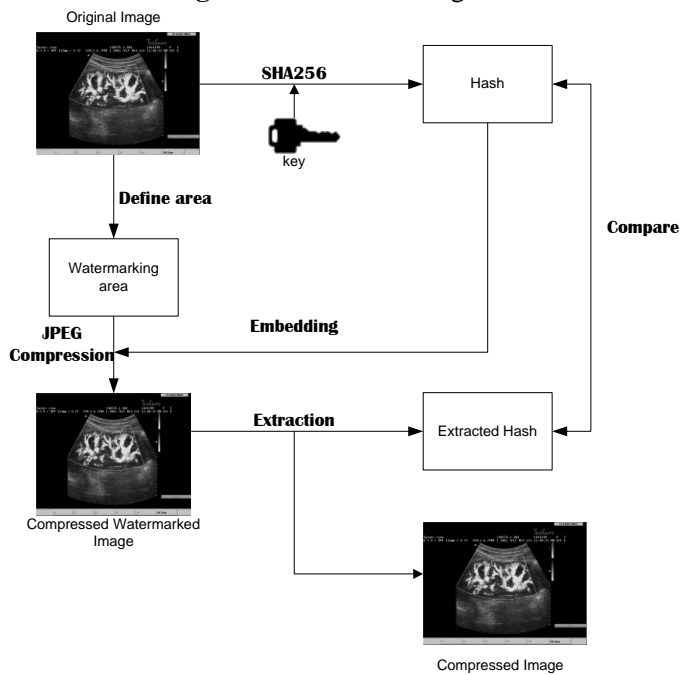

## 5. The Results

An ultrasound image of 800x600x8 with a watermark embedded in RONI was subject to increasing levels of JPEG compression. The compression was performed using a quality factor in Matlab 6.5.1 to produce files with .jpg extension.

The following are the results. Table 4 shows that the watermark is robust to a high compression rate up to 90.6%. The JPEG image quality threshold is 60 for the least significant bit embedding. The image quality threshold is increased to 61 for 2nd and 3rd LSB manipulations.

**Table 4**: LSB Embedding and Image Quality Threshold

| Manipulation | Image Quality Threshold | Compression (%) | PSNR (dB) |
|---|---|---|---|
| $1^{st}$ LSB | 60 | 90.6 | 40.75 |
| $2^{nd}$ LSB | 61 | 90.4 | 40.84 |
| $3^{rd}$ LSB | 61 | 90.4 | 40.84 |

Figure 10 shows the original 800x600 US image and the compressed watermarked image with quality 60. This has the effect of changing some pixel values, with a marked effect on areas of abrupt change resulting in the increase of pixel values 2 – 10 (figure 11). The effect of adding the watermark is evident by the peaks of pixel value 0 and 1. JPEG loses definition, particularly at high frequencies, which has the effect of low pass or smoothing filter.

An 800x600 ultrasound image was watermarked with its hash and then compressed with quality 60 in Matlab 6.5.1. The hash value of the original was recorded as "fcc29cbb8ea81be407cdd93e0326bf2bb68dca3d7872c9b6a033a981e184f989", and the hash value of the compressed image was extracted. Figure 4.19 shows (a) the original 800x600 ultrasound image, (b) the watermarked original image and (c) the watermarked image after compression with quality 60. The extracted hash value was "fcc29cbb8ea81be407cdd93e0326bf2bb68dca3d7872c9b6a033a981e184f989

", and is exactly the same as the original hash. This shows that the watermark can survive JPEG compression with Matlab 6.5.1 quality 60.

**Figure 10**: a) Original 800x600 US image b) compressed watermarked image with quality 60

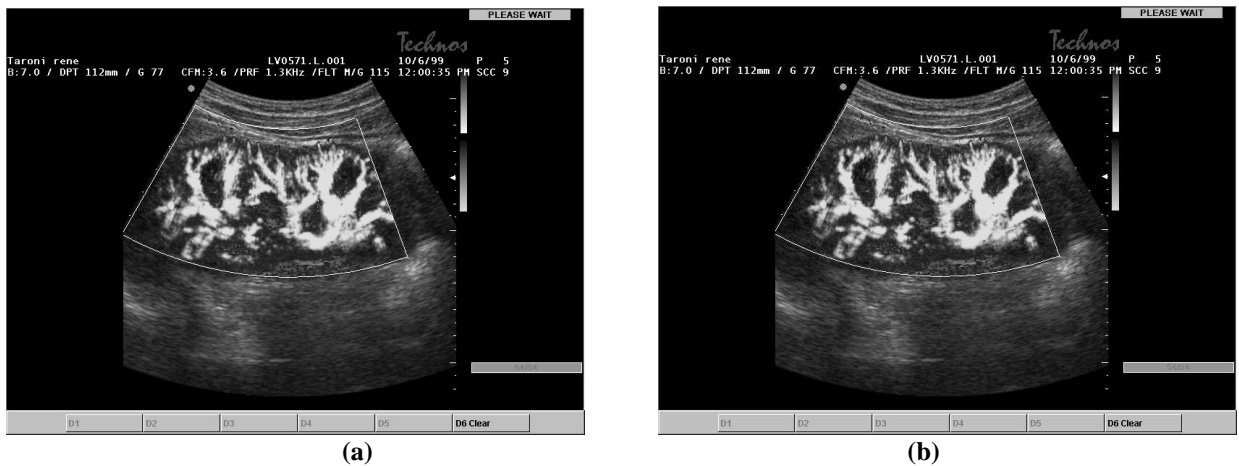

**Figure 11**: .a) Image histogram of figure 10(a); b) Image histogram of compressed image of figure 10(b)

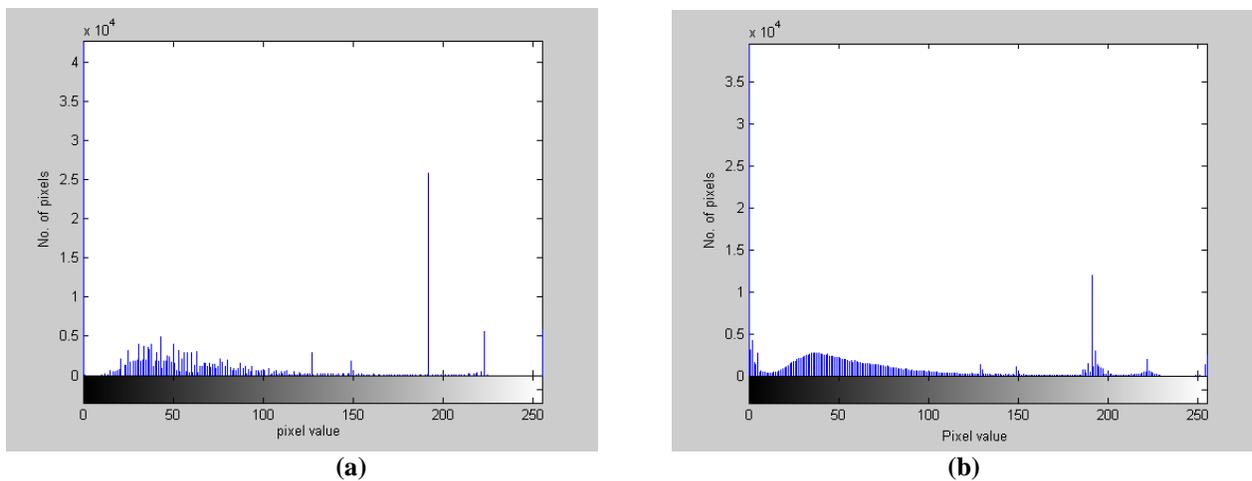

**Figure 12**: (a) original image (b) Watermarked image ( c) the image after compression

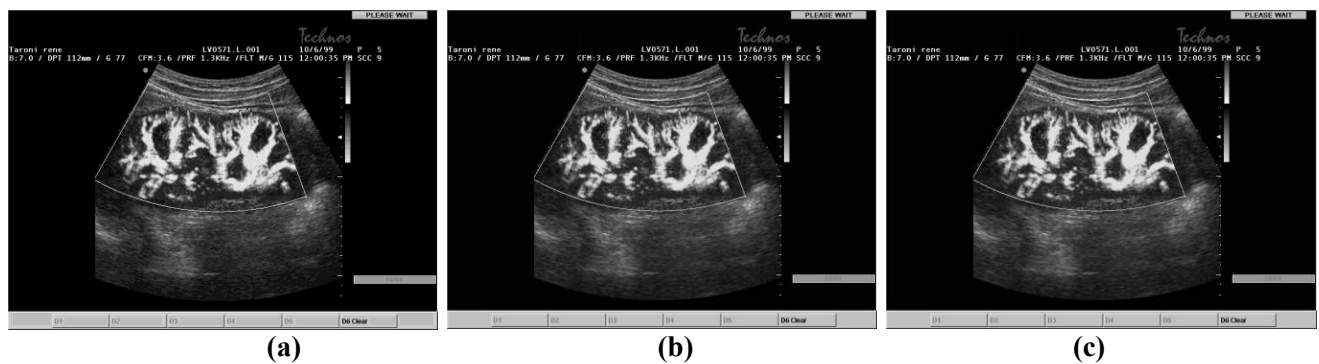

## 6. Summary and Concluding Remarks

A lossless watermarking scheme is proposed that is robust to lossy JPEG compression and at the same time is able to verify the authenticity and integrity of medical images. The watermarking scheme, including data embedding, extracting and verifying procedure were presented. Experimental results showed that such a scheme could embed and extract the watermark at a high compression rate. Combining cryptography and compression will add security to the medical images. In keeping the distortion level low, we could make sure that the watermarked image can still be valuable for other purposes, such as case studies in schools, but without disclosing a patient's confidential information.